# An oxorhenium complex bearing a chiral cyclohexane-1-olato-2-thiolato ligand: synthesis, stereochemistry and theoretical study of parity violation vibrational frequency shifts

Nidal Saleh,[a] Radovan Bast,*[b] Nicolas Vanthuyne,[c] Christian Roussel,[c] Trond Saue,[d] Benoît Darquié,*[e] and Jeanne Crassous*[a]

**Abstract:** In our effort towards measuring the parity violation energy difference between two enantiomers, a simple chiral oxorhenium complex **5** bearing enantiopure 2-mercaptocyclohexan-1-ol has been prepared as a potential candidate species. Vibrational circular dichroism revealed a chiral environment surrounding the rhenium atom, even though the rhenium is not a stereogenic centre itself, and enabled to assign the (1$S$,2$S$)-(-) and (1$R$,2$R$)-(+) absolute configuration for **5**. For both compound **5** and complex **4**, previously studied by us and bearing a propane-2-olato-3-thiolato ligand, relativistic calculations predict parity violating vibrational frequency differences of a few hundreds of millihertz, above the expected sensitivity attainable by a molecular beam Ramsey interferometer that we are constructing.

## Introduction

Chiral transition metal complexes have been a subject of interest in many fields such as asymmetric catalysis or molecular materials science.[1-4] They are currently considered as promising candidate molecules for the observation of parity violation (PV) effects.[5,6] Parity violating electroweak interactions break the mirror symmetry (under which all particle positions are reflected about a plane). This has a remarkable implication for molecular systems, as it should lead to a tiny energy difference between enantiomers of chiral molecules, which cease to be exact mirror images of each other and become diastereoisomers.[7-9] To date, no experiment has reached the energy resolution needed to observe the parity violation energy difference (PVED) between right- and left-handed isomers. Compared to the typical chiral prototype molecule CHFClBr (**1**, Figure 1), for which the equilibrium structure PVED is predicted to be around 30 mHz,[10,11] chiral heavy metal complexes are expected to show notably higher PVEDs of up to 500 Hz.[12-14] In this context, chiral oxorhenium complexes have attracted particular attention in the last decade. Several types of chiral oxorhenium(V) complexes have been prepared in enantiopure forms, such as complexes **2** and **3** (Figure 1) bearing respectively a chiral propane-2,3-diolato[15] and a dissymmetric sulfurated[16,17] ligand, both ligands inducing chirality at the rhenium centre. The stereochemistry of those complexes, revealed by vibrational circular dichroism (VCD), has been investigated in detail.[15-17]

Methyltrioxorhenium (MTO) derivatives substituted with a chiral ligand offer another appealing route to find candidate species for observing PV. One such complex bearing an enantiopure propane-2-olato-3-thiolato ligand[18] (complex **4** in Figure 1) has already been studied by us. In this paper, we report the successful preparation of a new enantiopure stable complex **5** (see Figure 1) bearing a cyclohexane-1-olato-2-thiolato ligand. We compare experimental and simulated VCD spectra, which we use for stereochemical characterization and for the determination of the absolute configuration. We furthermore carry out relativistic quantum chemical calculations of the PVED and the resulting vibrational frequency shift of the antisymmetric and symmetric Re=O stretching modes for both complexes **4** and **5**. Finally, we give a brief review of our progress towards measuring PV effects in heavy-metal chiral complexes using precise vibrational spectroscopy.

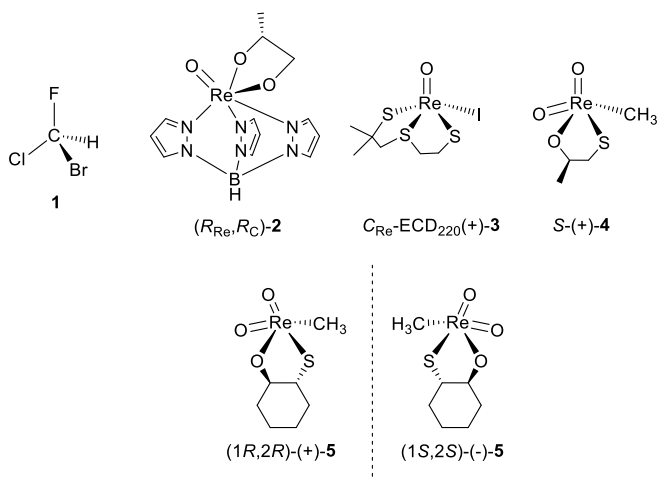

[a] Dr. N. Saleh, Dr. J. Crassous, Institut des Sciences Chimiques de Rennes UMR 6226, CNRS Université de Rennes 1, Campus de Beaulieu, 35042 Rennes Cedex (France). E-mail: jeanne.crassous@univ-rennes1.fr
[b] Dr. R. Bast, High Performance Computing Group, UiT The Arctic University of Norway, Tromsø, Norway. E-mail: radovan.bast@uit.no
[c] Dr. N. Vanthuyne, Prof. Christian Roussel, Aix Marseille Université, Centrale Marseille, CNRS, iSm2 UMR 7313, 13397, Marseille (France)
[d] Dr. T. Saue, Laboratoire de Chimie et Physique Quantiques, UMR 5626, CNRS et Université de Toulouse 3 (Paul Sabatier), 118 route de Narbonne, F-31062, Toulouse, France
[e] Dr B. Darquié, Laboratoire de Physique des Lasers, Université Paris 13, Sorbonne Paris Cité CNRS, F-93430, Villetaneuse, France. Email: benoit.darquie@univ-paris13.fr

**FIGURE 1** Chemical structures of chiral molecules synthesized for measuring parity violation effects.

**Materials and Methods**

Most experiments were performed using standard Schlenk techniques. Solvents were freshly distilled under argon from sodium/benzophenone (THF) or from phosphorus pentoxide ($CH_2Cl_2$). Starting materials were purchased from ACBR (MTO) or from Sigma Aldrich. Column chromatography purifications were performed in air over silica gel (Merck Geduran 60, 0.063–0.200 mm). $^1$H and $^{13}$C nuclear magnetic resonance (NMR) spectra were recorded on a Bruker AM 300 and 400. Chemical shifts were reported in parts per million (ppm) relative to $Si(CH_3)_4$ as external standard and compared to the literature. Infrared (IR) and VCD spectra were recorded on a Jasco FSV-6000 spectrometer. Thermogravimetric analysis (TGA) and differential scanning calorimetry (DSC) were carried out in flowing dry nitrogen, using a TGA/DSC 1 STARe System (METTLER TOLEDO) instrument. Specific rotations (in deg cm$^2$ g$^{-1}$) were measured in a 10 cm thermostated quartz cell on a Jasco P1010 polarimeter. Elemental analyses were performed by the CRMPO (*Centre régional de mesures physiques de l'Ouest*), University of Rennes 1.

**2-(Tritylthio)cyclohexanol 7**

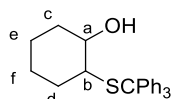

*n*-BuLi (10.5 mmol, 2.5 M, 4.2 mL) was added dropwise to a triphenylmethanethiol solution (9.9 mmol; 2.73 g) in distilled THF (20 mL) cooled to 0°C. Then cyclohexene oxide **6** (9.9 mmol, 1 mL) was added dropwise at 0°C with a color change from red to pale yellow. The reaction mixture was stirred for 24 hrs, then quenched with 20% AcOH in methanol, diluted with water, and then extracted with ethyl acetate. Purification over silica gel column chromatography (pentane/ethyl acetate; 9:1) provided the product **7** as a white precipitate (3.337 g, 90%). $^1$H NMR (400 MHz, CDCl$_3$, δ) 7.48 - 7.68 (6H, m, H$_{ar}$), 7.11 - 7.39 (9H, m, H$_{ar}$), 3.29 (1H, td, *J* = 9.1, 3.9 Hz, H$^a$), 2.07 - 2.21 (1H, m, H$^b$), 1.94 - 2.07 (1H, m, H$^c$), 1.58 (1 H, dd, *J* = 9.2, 3.64 Hz, H$^c$), 1.40 - 1.52 (2H, m, H$^d$), 0.83 - 1.36 (4H, m, H$^{e,f}$). Anal. calcd. for C$_{25}$H$_{26}$OS: C 80.17, H 7.00; found: C 80.15, H 7.09.

**2,2'-Disulfanediyldicyclohexanol 8**

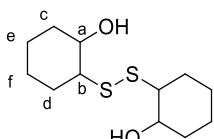

To a solution of 2-(tritylthio)cyclohexanol **7** (1 g, 2.67 mmol) in $CH_2Cl_2$/MeOH (9:1, 100 mL) was added in portions (over 30 min) an iodine solution (1.7 g in 150 mL of $CH_2Cl_2$/MeOH (9:1)). The reaction mixture was stirred at room temperature for 1 hr, then quenched with 10% aqueous sodium thiosulfate and washed with brine. The organic layer was separated and the aqueous layer was extracted with ethyl acetate, dried over MgSO$_4$, and concentrated under vacuum to provide a dark brown precipitate which was purified by silica gel chromatography (ethanol 5%/chloroform) to provide **8** as yellow-brown oil (350 mg, 90%). High-performance liquid chromatography (HPLC) resolution over chiral stationary phase enabled the separation of the two enantiomers (1*R*,1'*R*,2*R*,2'*R*)-(-)-**8** and (1*S*,1'*S*,2*S*,2'*S*)-(+)-**8**.$^1$H NMR (400 MHz, CDCl$_3$, δ) 3.53 (1H, td, *J* = 9.6, 4.6 Hz, H$^a$), 2.86 (1H, bs, OH), 2.52 - 2.67 (1H, m, H$^b$), 2.03 - 2.21 (2H, m, H$^c$), 1.71 - 1.86 (2 H, m, H$^d$), 1.21 - 1.60 (4 H, m, H$^{e,f}$). $^{13}$C NMR (101 MHz, CDCl3, δ) 72.8 (CH$^a$), 58.5 (CH$^b$), 34.2 (CH$_2^c$), 31.6 (CH$_2^d$), 26.1 (CH$_2^e$), 24.3 (CH$_2^f$). Anal. calcd. for C$_{12}$H$_{22}$O$_2$S$_2$: C 54.92, H 8.45; found: C 55.02, H 8.46.

The enantiomers showed mirror-image specific rotations $[\alpha]_D^{23}$ = -335/+333 (*C* = 3.8 x 10$^{-3}$ M, CH$_2$Cl$_2$).

**2-Mercaptocyclohexanol (1*S*,2*S*)-(+)-9 and (1*R*,2*R*)-(-)-9**[19]

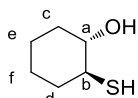

A solution of compound (1*S*,1'*S*,2*S*,2'*S*)-(+)-**8** (340 mg, 1.29 mmol) in THF (15 mL) was added dropwise to a stirring solution of LiAlH$_4$ (1.5 M, 3 eq.) in THF at 0°C. After stirring for 24 hrs at 50°C under argon, the reaction mixture was quenched with diluted HCl and then extracted with diethylether, dried over MgSO$_4$, and concentrated under vacuum to obtain (1*S*,2*S*)-(+)-**9** (280 mg, 86%) as a white solid. $[\alpha]_D^{23}$ = + 102 (*C* = 3.8 x 10$^{-3}$ M, CH$_2$Cl$_2$). $^1$H NMR (400 MHz, CDCl$_3$, δ) ppm 3.13 (1H, td, *J* = 9.85, 4.14 Hz, H$^a$), 2.63 (1H, bs, OH), 2.37 - 2.52 (1H, m, H$^b$), 1.95 - 2.15 (2H, m, H$^c$), 1.72 (1H, dd, J = 9.3, 2.8 Hz, H$^d$), 1.59 - 1.68 (1 H, m, H$^d$), 1.15 - 1.39 (5H, m, SH, H$^{e,f}$). $^{13}$C NMR (101 MHz, CDCl$_3$, δ) 76.65 (CH$^a$), 47.68 (CH$^b$), 36.50 (CH$_2^c$), 34.03 (CH$_2^d$), 26.54 (CH$_2^e$), 24.72 (CH$_2^f$). Anal. calcd. for C$_6$H$_{12}$OS: C 54.50, H 9.15; found: C 54.42, H 9.29.
(1*R*,2*R*)-(-)-**9** was prepared using the same procedure but starting with (1*R*,1'*R*,2*R*,2'*R*)-(-)-**8**: $[\alpha]_D^{23}$ = - 98 ( *C* = 3.8 x 10$^{-3}$ M, CH$_2$Cl$_2$).

**Oxorhenium complex (1*S*,2*S*)-(-)-5 and (1*R*,2*R*)-(+)-5**

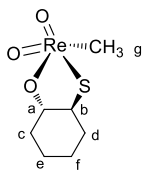

To a solution of MTO (94 mg, 0.37 mmol) in CH$_2$Cl$_2$ (7 mL), was added a solution of (1$S$,2$S$)-(+)-2-mercaptocyclohexanol **9** (50 mg, 0.37 mmol) in CH$_2$Cl$_2$ (3 mL). After stirring for 3 hrs under argon at room temperature, the solvent was stripped off, yielding quantitatively (1$S$,2$S$)-(-)-**5** (134 mg) as an orange-red precipitate. $[\alpha]_D^{23}$ = -86 (C = 2.75 x 10$^{-3}$ M, CH$_2$Cl$_2$). $^1$H NMR (400 MHz, CDCl$_3$, δ) 3.90 (1H, td, $J$ = 10.7, 3.9 Hz, H$^a$), 3.19 - 3.32 (1H, m, H$^b$), 2.40 (3H, s, H$^g$), 2.05 - 2.19 (2H, m, H$^c$), 1.65 - 1.78 (2H, m, H$^d$), 1.31 - 1.50 (2H, m, H$^e$), 1.08 - 1.26 (2H, m, H$^f$). $^{13}$C NMR (101 MHz, CDCl$_3$, δ) 90.46 (CH$^a$), 57.75 (CH$^b$), 33.67 (CH$_2$$^c$), 32.62 (CH$_2$$^d$), 30.16 (CH$_3$$^g$), 24.54 (CH$_2$$^e$), 22.69 (CH$_2$$^f$). Anal. calcd. for C$_7$H$_{13}$O$_3$ReS: C 23.13, H 3.61; found: C 22.02, H 3.70.

The same procedure was used for the preparation of the other enantiomer (1$R$,2$R$)-(+)-**5**: $[\alpha]_D^{23}$ = +84 ($C$ = 2.75 x 10$^{-3}$ M, CH$_2$Cl$_2$).

**Chiral HPLC separation of compound 8**

The sample is dissolved in ethanol, injected on the chiral columns, and detected with an UV detector at 254 nm and a polarimeter. The flow-rate is 1 mL/min. Major products are *meso*-**8**, (-)-**8** and (+)-**8**, sign given by the on-line polarimeter in the mobile phase used. See SI for more details.

**Computational details**

The equilibrium molecular structures and the harmonic force fields were solved with the Gaussian 09 package[20] using Kohn-Sham density functional theory with the B3LYP functional[21,22] and Ahlrichs' def2-TZVPP basis sets.[23,24] The PV vibrational frequency shifts were obtained by computing PV expectation values using single-point calculations along the normal mode coordinate displacement vectors (single-point structures are listed in the SI). The single-point 4-component relativistic calculations using the Dirac/Coulomb Hamiltonian were performed with the DIRAC program[25] using the HF method, as well as the functionals B3LYP and PBE[26]. These provided the anharmonic potential for the numerical solution of the vibrational wave functions by the Numerov-Cooley algorithm. The Numerov-Cooley solutions and integrations were performed using a Python script that we have written for this project.[27]

**Results and Discussion**

**Synthesis of enantiopure 2-mercaptocyclohexan-1-ol**

Racemic 2-mercaptocyclohexan-1-ol (±)-**9** can be readily prepared from cyclohexene epoxide **6** in three steps reactions as summarized in Scheme 1. This synthetic strategy is inspired by the literature and by our previous work regarding the preparation of enantiopure 2-methyl-thioethanol from propylene oxide enantiomers.[18] First, the opening of commercially available achiral cyclohexene epoxide **6** with trityl thiol in the presence of *n*-BuLi gave the alcohol **7** with 90% yield. This step introduces the two asymmetric carbons. Then the deprotection followed by the *in situ* oxidation with iodine yielded the disulfure compound **8** as a statistical mixture of racemic and *meso* compounds with 90% yield. This mixture was finally reduced by LiAlH$_4$ to give 2-mercaptocyclohexan-1-ol **9** with 86% yield.

Chiral HPLC over a chiral stationary phase was used to separate the disulfide intermediate **8** and enabled to isolate the two enantiomers (1$S$,1'$S$,2$S$,2'$S$)-(+)-**8** and (1$R$,1'$R$,2$R$,2'R)-(-)-**8** from the *meso*-**8** compound (See SI). Then, the reduction of enantiopure disulfide **8** by LiAlH$_4$ yielded (1$S$,2$S$)-(+)-**9** and (1$R$,2$R$)-(-)-**9** enantiomers. The specific rotation values $[\alpha]_D^{23}$ (Table 1) support the mirror-image relationship within the experimental errors (±2%).

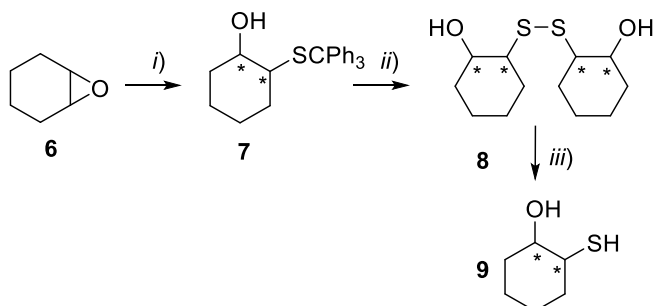

**SCHEME 1** Synthetic pathway of racemic 2-mercaptocyclohexan-1-ol (±)-**9**. *i*) *n*-BuLi, Ph$_3$CSH, THF, 0 °C, 24 hrs, 90%; *ii*) I$_2$, CH$_2$Cl$_2$, MeOH, 30 min, 90%; *iii*) LiAlH$_4$, THF, 50°C, Ar, 24 hrs, 86%.

**Synthesis of enantiopure oxorhenium complexes**

Finally, enantiopure oxorhenium complexes (+) and (-)-**5** were prepared by simply mixing MTO and enantiopure (-)- and (+)-2-mercaptocyclohexanol **9**, respectively, in anhydrous CH$_2$Cl$_2$ at room temperature for 3 hours (Scheme 2). After vacuum removal of CH$_2$Cl$_2$, an easily handled red powder was observed corresponding to **5**, as verified by $^1$H NMR that showed deshielding effect of the protons' resonances of **9** after complexation. Moreover, the complexes show a good stability in the solid state and in solution and possess the ability to sublime at 40-60 °C under reduced pressure (~10$^{-2}$ Pa).

Importantly, the presence of enantiopure ligand provides a chiral environment around the rhenium atom, although the rhenium is not a stereogenic center in itself. This chiral environment is confirmed by measuring the specific rotation values $[\alpha]_D^{23}$ for (1$R$,2$R$)-(+)-**5** and (1$S$,2$S$)-(-)-**5** (+84 and -86, respectively; C = 2.75 x 10$^{-3}$ M in CH$_2$Cl$_2$) and also by VCD spectroscopy data (*vide infra*).

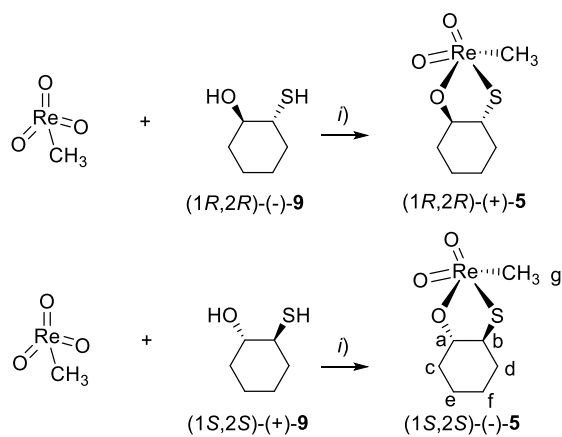

**SCHEME 2.** Synthesis of the enantiopure oxorhenium complexes (1*S*,2*S*)-(-)-**5** and (1*R*,2*R*)-(+)-**5** from MTO and enantiopure **9**. *i*) CH$_2$Cl$_2$, rt, 3 hrs, quant.

**TABLE 1** Experimental specific rotations values (*C* = 3.5-4 x 10$^{-3}$ M in CH$_2$Cl$_2$)

|  | (1*S*,1'*S*,2*S*,2'*S*)-(+)-**8** | (1*R*,1'*R*,2*R*,2'*R*)-(-)-**8** | 1*S*,2*S*-(+)-**9** | 1*R*,2*R*-(-)-**9** |
|---|---|---|---|---|
| $[\alpha]_D^{23}$ | +333 | -335 | +102 | -98 |

**VCD spectroscopy**

Vibrational circular dichroism is a chiroptical technique enabling to determine the absolute configuration, through comparison between experimental measurements and theoretical calculations.[28-30] In addition, VCD can be a very powerful tool to study the dissymmetric environment around the chiral metal center.[31-33] As seen in Figure 2, a) and b), IR spectra of the two enantiomers measured in CD$_2$Cl$_2$ coincide properly and display several bands, such as the characteristic Re=O stretching band at ~1030 cm$^{-1}$. The experimental VCD spectra of (+)- and (-)-**5** in CD$_2$Cl$_2$ show an overall mirror-image relationship (Figure 2, c) and d)), with, for (1*R*,2*R*)-(+)-**5** in red, strong and consecutive positive-negative VCD bands at 1344-1312 cm$^{-1}$ respectively, and negative-positive VCD bands at 1026-1021 cm$^{-1}$, respectively. The band at 1026 cm$^{-1}$ may be tentatively assigned to the Re=O stretching band and displays a dissymmetry factor $\Delta\varepsilon/\varepsilon$ = 1.1x10$^{-4}$. Interestingly, the presence of such a VCD active band indicates a chiral environment around the rhenium atom, although the Re in itself is not a stereogenic element.

The IR and VCD spectra of complex **5** were calculated and compared to the experimental data (see VCD spectra comparison in Figure 3). First, the structure having the (1*R*,2*R*) stereochemistry was optimized at the B3LYP / def2-TZVPP[23,24] level of theory. Only one stable conformer, displayed in Figure 3 and above Table 2, was obtained, which corresponds to the cyclohexane ring adopting a chair conformation, with the S and O3 atoms placed in the equatorial positions. Moreover, the rhenium center is pentacoordinated, the C1 atom is *trans* to the O3 oxygen, and the Re, O3, S, C1 lie in the same plane, while the two oxo groups are symmetrically placed on each side of this plane. Consequently, the oxygen atoms O1 and O2 only differ in their chemical environment due to the proximity of the two asymmetric carbons. This structure is similar to the chiral oxorhenium complex **4** previously described by us.[18] The VCD spectrum of (1*R*,2*R*)-**5** was then calculated by TDDFT at the same B3LYP / def2-TZVPP[23,24] level of theory and appeared to agree very well with the experimental one (see Figure 3), thus enabling to determine the absolute configurations (1*S*,2*S*)-(-) and (1*R*,2*R*)-(+) of the oxorhenium complex **5**, and to deduce the (1*S*,2*S*)-(+) and (1*R*,2*R*)-(-) absolute configurations for free ligand **9**, and (1S,1'*S*,2*S*,2'*S*)-(+)-**8** and (1*R*,1'*R*,2*R*,2'*R*)-(-)-**8** for the corresponding disulfide.

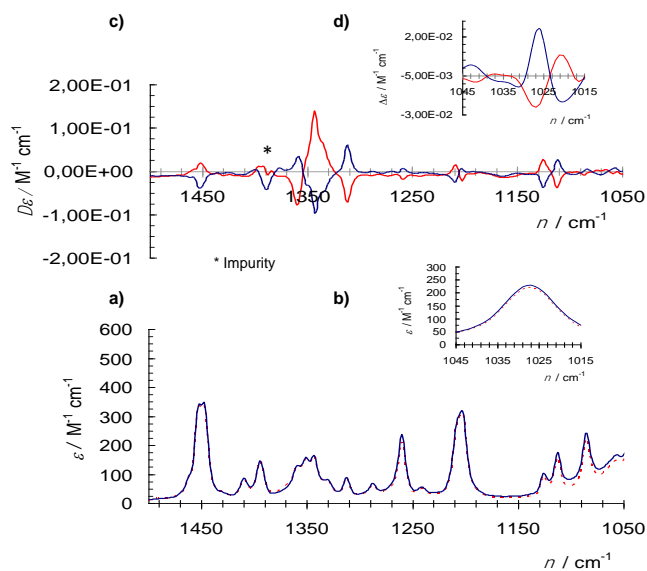

**FIGURE 2.** Experimental a, b) IR and c, d) VCD spectra of (+)-**5** (red curve) and (-)-5 (blue curve) (C = 0.055 and 0.011 M in CD$_2$Cl$_2$, 500 μm path length cells with BaF$_2$ windows).

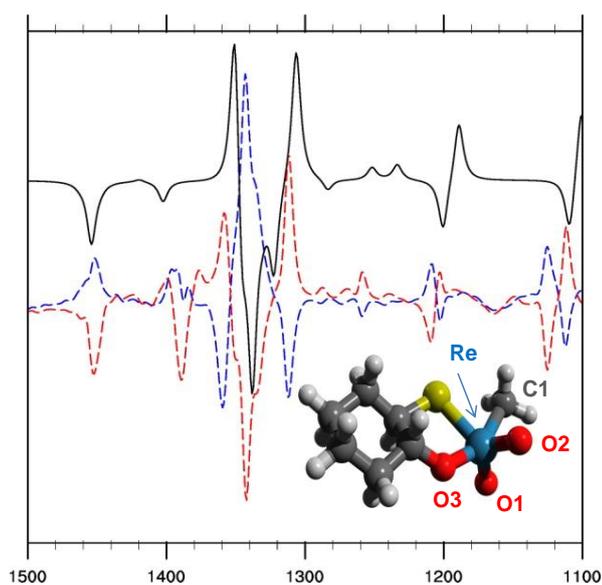

**FIGURE 3.** Overlay of the simulated VCD spectrum for complex (*R,R*)-**5** (black solid line, B3LYP, scaled), and the experimental spectra (red and blue dashed lines corresponding respectively to (+)-**5** and (-)-**5**). The B3LYP (def2-TZVPP basis) equilibrium structure of the theoretically studied stable conformer is also displayed.

**TGA/DSC Analysis**

To evaluate the thermal stability of complex **5**, thermogravimetric analysis (TGA) and differential scanning calorimetry (DSC) were carried out on a ~6 mg sample, at a heating rate of 5°C/min. The thermogram shown in Figure 4 indicates that complex **5** is thermally stable up to nearly 136 °C. The DSC curve shows an endothermic peak at 44 °C without mass loss in the TGA curve. This result is in agreement with the sublimation results (*vide supra*) and confirms that complex **5** may be suitable for future PV measurements based on molecular beam spectroscopy experiments. It therefore corresponds to its melting point. At 136°C, an endothermic peak is observed in the DSC curve, associated to a mass loss in the TGA curve. This is the signature of thermal decomposition with the formation of volatile products.

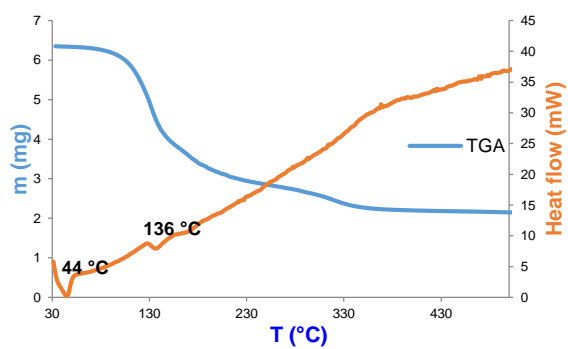

**FIGURE 4.** TGA (blue curve, left axis) – DSC (orange curve, right axis) of oxorhenium complex **5**. Carried out in flowing dry nitrogen, ~6 mg sample, 5°C/min heating rate.

## PV calculations

Relativistic calculations of PV vibrational frequency shifts of the symmetric and antisymmetric Re=O stretching modes were performed for both the new complex **5** and complex **4** previously studied by us[18]. Those are summarized in Table 2. Compound **4** has two conformations **4-c1** and **4-c2**, the calculated equilibrium structures of which are displayed above Table 2. For the two conformers of **4** the results obtained in our earlier work[18] are reported together with new investigations. In our previous study we used the same method to obtain the harmonic force field (marked as hFF in Table 2), the anharmonic contributions along the normal mode displacement (marked as aFF), as well as to compute the energy and the PVED along a selected normal mode coordinate. In this work we have varied methods/functionals to obtain 1) the normal mode coordinate displacements, 2) to compute the energy, and 3) the PVED along the normal mode coordinate displacements in order to assess the sensitivity of these contributions. We report values on the order of 0.1-0.2 Hz for complex **5** and up to 0.8 Hz for complex **4**, confirming earlier studies on such compounds. We observe that the computed PV vibrational frequency shifts vary significantly with the employed method, even changing the sign, with the HF results being generally larger in magnitude compared to B3LYP and PBE. The change is less pronounced when varying the method for computing the anharmonic contributions to the potential while keeping the harmonic force field method fixed. Changing the aFF method has a smaller effect for the antisymmetric stretch compared to the symmetric Re=O stretching mode. This can be rationalized by comparing the harmonic frequencies with the fundamental transition frequencies obtained by the Numerov-Cooley method: the antisymmetric stretch seems to be less anharmonic than the symmetric stretch. In the latter case the harmonic frequency is a poorer approximation and at the same time the values vary more when modifying the aFF method. We conclude this discussion by reiterating that an accurate description of the density around the heavy center seems crucial for a reliable prediction of the expectation value and seems to have the most important effect.

**TABLE 2.** Computed harmonic and fundamental vibrational transition frequencies (in cm$^{-1}$) of the antisymmetric and symmetric Re=O stretching modes for the molecules/conformers **4-c1**, **4-c2**, (see reference 18 for details) and **5**, with corresponding 4-component DC Hamiltonian PV fundamental transition frequency differences. We distinguish between methods that yielded the harmonic force field (hFF), anharmonic force field (aFF) and methods used to compute expectation values along the normal mode coordinate displacements (PV). The B3LYP (def2-TZVPP basis) equilibrium structures of the species studied are also displayed.

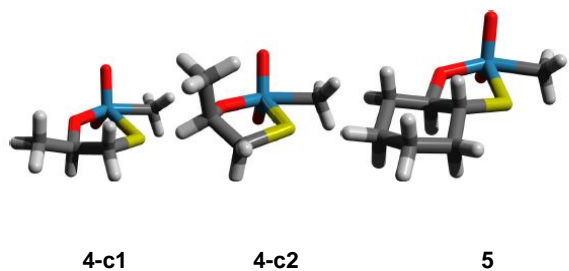

    **4-c1**        **4-c2**        **5**

| Mole-cule | Vib. mode | Freq. (cm$^{-1}$) | | Method | | | PV shift (Hz) |
|---|---|---|---|---|---|---|---|
| | | harm. | fund. | hFF | aFF | PV | |
| **4-c1** | asym | 986[a] | 982[a] | B3LYP | B3LYP | B3LYP | 0.078[a] |
| | | 986 | 982 | B3LYP | B3LYP | PBE | 0.166 |
| | | 986 | 980 | B3LYP | PBE | PBE | 0.163 |
| | | 951[a] | 944[a] | PBE | PBE | PBE | 0.195[a] |
| | | 986 | 982 | B3LYP | B3LYP | HF | -0.828 |
| | | 986 | 966 | B3LYP | HF | HF | -0.833 |
| | | 1106[a] | 1102[a] | HF | HF | HF | -0.211[a] |
| | sym | 1022[a] | 1010[a] | B3LYP | B3LYP | B3LYP | 0.150[a] |
| | | 1022 | 1010 | B3LYP | B3LYP | PBE | 0.019 |
| | | 1022 | 969 | B3LYP | PBE | PBE | 0.055 |
| | | 980[a] | 968[a] | PBE | PBE | PBE | 0.036[a] |
| | | 1022 | 1010 | B3LYP | B3LYP | HF | 0.478 |
| | | 1022 | 1160 | B3LYP | HF | HF | 0.052 |
| | | 1174[a] | 1165[a] | HF | HF | HF | 0.344[a] |
| 4-c2 | asym | 985[a] | 981[a] | B3LYP | B3LYP | B3LYP | -0.119[a] |
| | | 985 | 981 | B3LYP | B3LYP | PBE | -0.237 |
| | | 985 | 979 | B3LYP | PBE | PBE | -0.234 |
| | | 950[a] | 944[a] | PBE | PBE | PBE | -0.271[a] |
| | | 985 | 981 | B3LYP | B3LYP | HF | 0.836 |
| | | 985 | 966 | B3LYP | HF | HF | 0.839 |
| | | 1106[a] | 1102[a] | HF | HF | HF | 0.219[a] |
| | sym | 1021[a] | 1010[a] | B3LYP | B3LYP | B3LYP | -0.170[a] |
| | | 1021 | 1010 | B3LYP | B3LYP | PBE | -0.049 |
| | | 1021 | 974 | B3LYP | PBE | PBE | -0.086 |
| | | 979[a] | 966[a] | PBE | PBE | PBE | -0.084[a] |
| | | 1021 | 1010 | B3LYP | B3LYP | HF | -0.708 |
| | | 1021 | 1138 | B3LYP | HF | HF | -0.182 |
| | | 1175[a] | 1165[a] | HF | HF | HF | -0.056[a] |
| 5 | asym | 985 | 981 | B3LYP | B3LYP | B3LYP | 0.035 |
| | sym | 1022 | 1010 | B3LYP | B3LYP | B3LYP | 0.125 |

[a] See ref. 18.

**Toward measuring parity violation vibrational frequency differences between enantiomers of complex 5**

We briefly summarize here the main results that have been obtained by our consortium toward measuring PV effects in chiral molecules by ultra-high resolution spectroscopy. Our approach is based on the 1975 suggestion by Letokhov to search for a shift $\Delta\nu_{PV} = \nu_L - \nu_R$ in the frequencies $\nu_L$ and $\nu_R$ of the same rovibrational transition of left and right enantiomers,[34] associated with the PVED. Around 2000, the Laboratoire de Physique des Lasers, carried out such high-resolution experiments by probing a hyperfine component of CHFClBr **1** molecule's C-F stretch (see Figure 1) of frequency $\nu \sim 30$ THz (~10 µm), using saturated absorption laser spectroscopy.[35,36] A world-record experimental sensitivity of $2 \times 10^{-13}$, corresponding to $\Delta\nu_{PV} < 8$ Hz was demonstrated, limited by residual differential pressure shifts induced by impurities in the samples. This work then triggered theoretical studies which concluded that the PV shift on vibrational transitions of CHFClBr are in fact very small (~2 mHz[10,11]), but that chiral complexes of heavy metals studied by Schwerdtfeger, Bast and coworkers, have predicted PV shifts as large as 1 Hz.[12,13,37] To be able to measure $\Delta\nu_{PV}$, we must thus *i*) improve the sensitivity of the measurement by at least an order of magnitude and *ii*) work with heavy-metal containing species.

For the former, the proposed improvement is to perform 2-photon Ramsey interferometry on a molecular beam using ultra-stable lasers. This is expected to allow PV shifts to be evaluated to below 1 part in $10^{15}$ (a few tens of millihertz for transitions at ~30 THz).[6,38,39] For the production of better molecular samples, our consortium has focused on chiral rhenium complexes. This led to the successful production of several classes of oxorhenium compounds in enantiopure form such as complexes **2**, **3**, **4** and **5**. Our calculations (references 17,18 and this work) indicate that PV shifts in species such as **3**, **4** and **5** can reach several hundreds of millihertz, which is above the expected sensitivity attainable by Ramsey interferometry on a molecular beam.[6,38,39] Compared to **3** and **4**, species **5** shows an improved thermal stability and is as such particularly promising.

To gain insight in the apparatus and know-how required for performing precise spectroscopic measurements on such complex species, we have conducted high resolution mid-infrared spectroscopy of the achiral precursor MTO **1**, a model achiral complex that can be very easy sublimed, in both room temperature cells and cold supersonic beams using frequency stabilised $CO_2$ lasers or quantum cascade lasers[39-42]. We have also demonstrated buffer-gas-cooling of laser-ablated gas phase MTO[42,43] and precise spectroscopy in a 6 K helium buffer gas cell. This shows that organo-metallic species of interest for a PV measurement survive laser ablation and are thus well-suited to the generation of buffer-gas molecular beams, which, compared to their supersonic counterpart, generally exhibit higher fluxes and lower translational velocities, thus leading to improved resolutions. Those measurements have now to be repeated on chiral candidates and in particular on complex **5**. On the laser side, we have recently demonstrated record frequency stabilities and accuracies necessary for performing Ramsey interferometry with both $CO_2$ lasers (typically used in this region for precise spectroscopy) and quantum cascade lasers.[41,44,45] Quantum cascade lasers cover the entire mid-infrared and are far more tuneable than $CO_2$ lasers, and will thus allow in the near future many more candidate chiral species to be studied.

Note that we have recently found a uranium compound (N≡UHFI) with a record ~20 Hz predicted PV frequency shift.[46] Although synthesizing and isolating such compounds has not been demonstrated so far, this could be possible in the future.

**Conclusion**

We have prepared a new chiral oxorhenium complex, (1*S*,2*S*)-(-) and (1*R*,2*R*)-(+)-**5**, which may be considered as a new candidate species for a parity violation measurement. Although the rhenium center is not chiral, its chiral environment has been characterized by VCD spectroscopy. For the last five years, we have carried out ultra-high resolution vibrational spectroscopy of rhenium complexes and believe that it may be used for PV measurements. Relativistic quantum chemical calculations predict PV vibrational frequency shifts of a few hundreds of millihertz for complexes such as **5**, above the sensitivity attainable via precise spectroscopic measurements. Progresses on both spectroscopic measurements and the synthesis of new chiral transition metal complexes are ongoing.

**Acknowledgements**

We thank the Ministère de l'Education Nationale, de la Recherche et de la Technologie, the Centre National de la Recherche Scientifique (CNRS) and the University of Rennes 1. This work was supported by ANR (under grants no ANR 2010 BLAN 724 3 and no ANR-15-CE30-0005-01) and Région Île-de-France (DIM Nano-K). The authors thank A Amy-Klein, P Asselin, C Chardonnet, C Daussy, L Guy, TR Huet, P Soulard, and SK Tokunaga for fruitful discussions.

**Supporting information**

Additional supporting information may be found in the online version of this article at the publisher's website: the experimental HPLC separations and Numerov script input and output data, plots, scripts, optimized structures, and displaced structure data for single-point calculations.